\PassOptionsToPackage{usenames,dvipsnames}{xcolor}
\documentclass[a4paper,11pt,2p]{elsarticle}

\usepackage[T1]{fontenc}
\usepackage[utf8]{inputenc}
\usepackage[british]{babel}
\addto\captionsbritish{}
\usepackage{slashbox} 
\usepackage{geometry}
\usepackage{graphicx} 
\usepackage{amssymb} 
\usepackage{amsmath,pdftexcmds}
\usepackage{fancyvrb}
\usepackage{multicol}
\usepackage{multirow}
\usepackage{mathtools}
\usepackage{bbm}
\usepackage{amsbsy}
\usepackage{tabularx}
\usepackage{hyperref}
\usepackage{adjustbox}  
\usepackage{url}
\usepackage{booktabs}
\usepackage{mathrsfs}
\usepackage{pgffor}
\usepackage{textcomp}
\usepackage{pgfplots}
\usepackage{pgfplotstable}
\pgfplotsset{compat=1.6}
\usepackage[version=3]{mhchem}
\usepackage{framed}
\usepackage{pgfplots}
\usepackage{comment}
\usepackage{youngtab}
\usepackage{tikz}
\usepackage{tikz-3dplot}
\usetikzlibrary{trees}
\usetikzlibrary{decorations.markings}
\usetikzlibrary{calc}
\usetikzlibrary{plotmarks}
\usepgfplotslibrary{fillbetween}
\usepackage[all]{xy}
\usepackage{amsfonts}
\usetikzlibrary{arrows}
\usepackage{anysize}
\marginsize{2.4cm}{2.4cm}{1.2cm}{2.6cm}

\DeclareFontFamily{U}{cbgreek}{}
\DeclareFontShape{U}{cbgreek}{m}{n}{
        <-6>    grmn0500
        <6-7>   grmn0600
        <7-8>   grmn0700
        <8-9>   grmn0800
        <9-10>  grmn0900
        <10-12> grmn1000
        <12-17> grmn1200
        <17->   grmn1728
      }{}
\DeclareFontShape{U}{cbgreek}{bx}{n}{
        <-6>    grxn0500
        <6-7>   grxn0600
        <7-8>   grxn0700
        <8-9>   grxn0800
        <9-10>  grxn0900
        <10-12> grxn1000
        <12-17> grxn1200
        <17->   grxn1728
      }{}

\DeclareRobustCommand{\Qoppa}{%
  \text{\usefont{U}{cbgreek}{\normalorbold}{n}\symbol{21}}%
}

\DeclareRobustCommand{\digamma}{%
  \text{\usefont{U}{cbgreek}{\normalorbold}{n}\symbol{147}}%
}

\makeatletter
\newcommand{\normalorbold}{%
  \ifnum\pdf@strcmp{\math@version}{bold}=\z@ bx\else m\fi
}
\makeatother

\makeatletter
\def\ps@pprintTitle{
   \let\@oddhead\@empty
   \let\@evenhead\@empty
   \def\@oddfoot{\reset@font\hfil\thepage\hfil}
   \let\@evenfoot\@oddfoot
}
\makeatother

\newsavebox{\leftbox}
\newsavebox{\rightbox}

\definecolor{DarkMidnightBlue}{rgb}{0.0, 0.04, 0.14}

\hypersetup{hidelinks,
backref=true,
pagebackref=true,
hyperindex=true,
breaklinks=true,
colorlinks=true,
urlcolor=DarkMidnightBlue,
bookmarks=true,
bookmarksopen=false,
pdftitle={Title},
pdfauthor={Author}}

\VerbatimFootnotes

\title{\textbf{\textsf{Two-Fermion Bound and Scattering States in a Finite Volume including QED in P-Wave and Beyond}}}

\author[hiskp,bctp]{Gianluca~Stellin}
\ead{stellin@hiskp.uni-bonn.de}

\address[hiskp]{Helmholtz Institut für Strahlen- und Kernphysik, Universität Bonn, Nu\ss{}allee 14-16, 53115 Bonn, Germany}
\address[bctp]{Bethe Center for Theoretical Physics, Universität Bonn, Nu\ss{}allee 12, 53115 Bonn, Germany}

\date{\today}

\begin{document}

\begin{abstract}
\begin{small}
Introducing a short range force coupling the spinless fermions to one unit of angular momentum in the framework of 
pionless EFT, we first report the two-body scattering amplitudes with Coulomb corrections, extended to two fermions of opposite 
charge in refs.~\cite{StM21,Ste20}. Motivated by the growing interest in lattice approaches, we immerse the system into a cubic 
box with periodic boundary conditions and display the finite-volume corrections to the energy of the lowest bound and unbound $T_1^{-}$ 
eigenstates. The latter turn out to consist of power law terms proportional to the fine-structure constant. In the calculations, quadratic and 
higher order contributions in $\alpha$ are discarded, on the grounds that the gapped nature of the momentum operator in the 
finite-volume environment allows for a perturbative treatment of the QED interactions. An outlook on the extension of the analysis 
to D-wave short-range interactions is eventually given.
\end{small}
\end{abstract}

\maketitle

\bigskip 

\bigskip

\bigskip

\tableofcontents

\bigskip 

\bigskip

\bigskip


\section{Preamble}\label{S-1.0}

In present times, effective field theories (EFT) \cite{LaM19}
cover a crucial role in the description of many-body systems 
in nuclear and subnuclear physics, availing of the quantum fields which can be excited 
in a given regime of energy. For systems of stable baryons at energies lower 
than the pion mass, the Lagrangian density is defined uniquely as a functional of the nucleon fields 
and their Hermitian conjugates. The resulting theory \cite{KSW96,KSW98-01}
numbers several successes in the context of nucleon-nucleon scattering and 
structure properties of few-nucleon systems.
In the latter, the Power Divergence Subtraction (PDS) is adopted 
as a regularization scheme \cite{KSW98-01,KSW98-02}.

Based on the p-wave interactions in the EFT for halo nuclei, we generalize 
in secs.~\ref{S-2.0} and \ref{S-2.1} the nonperturbative application of QED in pionless EFT
in ref.~\cite{KoR00} to fermion-fermion low-momentum elastic scattering governed by the 
the Coulomb and the strong forces transforming as the $\ell=1$ representation of the rotation 
group. In sec.~2.3 of ref.~\cite{StM21} we have applied the formalism also to 
fermion-antifermion scattering, where the attractive Coulomb force 
gives rise also to bound states, as in the case of the \textit{protonium} \cite{KBM02}. 

Aware of the importance of the lattice environment in present-day EFT and QCD calculations \cite{DaS14,BDH21}, in sec.~3 of 
ref.~\cite{StM21}, we transposed our fermion-fermion EFT into a cubic spatial volume with side $L$ and periodic boundary conditions 
(PBC). The finite-volume (FV) environment generates a number of implications, the most glaring of them are the breaking of rotational 
symmetry and the discretization of the spectrum of the operators representing \cite{SEM18} physical observables \cite{KLH12}.

By virtue of the long-range interactions induced by QED, significant changes in the behaviour as a function of $L$ of the 
corrections associated to the FV energy levels are observed. The latter assume the form of polynomials in
the reciprocal of the box size \cite{DaS14}, that take the place of the exponential damping factors encountered with short-range 
forces alone \cite{Lue86-01,Lue86-02,Lue91,KLH12}. 
Besides, the gapped nature of the momentum of the particles in the box allows for a perturbative treatment of the QED contributions,
even at low energies \cite{DaS14,BeS14}. Consequently, composite particles receive modifications of the same kind both in their 
mass \cite{DaS14} and in the energies of the two-body bound and unbound states they give rise to \cite{BeS14}.

In scattering states, the formula for the shift with respect to the infinite volume energy for the lowest P-wave state
in sec.~\ref{S-3.2.1} shows some similarities with the one in ref.~\cite{BeS14}, 
despite an overall $\xi/M \equiv 4\pi^2/M L^2$ factor, owing to the fact that the energy of the lowest unbound state 
with analogous symmetry properties under rotations ($T_1$ irrep of the cubic group) is nozero. 

Concerning the leading-order (LO) energy shift for the lowest bound state, in the P-wave case it is proportional
to $\alpha_{\mathrm{QED}} \equiv \alpha$ and has the same sign of the counterpart in absence of QED in
refs.~\cite{KLH12}. Additionally, we prove that the FV shifts for S- (cf. ref.~\cite{BeS14}) and P-wave
eigenenergies have the same magnitude if order $1/L^3$ terms are neglected, as in absence of QED. 

Even if bound states between two hadrons of the same charge have not been identified in nature, at unphysical values
of the quark masses in Lattice QCD these states begin to appear. It is not unlikely that
the latter manifest themselves also when QED becomes part of the Lagrangian. 


\section{Effective field theory for non-relativistic fermions}\label{S-2.0}

Our analysis of two-particle scattering and bound states in the infinite- and finite-volume context
hinges on pionless EFT \cite{KSW96,KSW98-01,KSW98-02}, describing many-nucleon systems at energies smaller 
than the pion mass, $M_{\pi}$ \cite{KoR00,LaM19}. The action is constructed on non-relativistic matter fields 
and is left invariant by parity, time reversal and Galilean invariance. 

We apply the framework to spinless fermions (e.g. baryons) of mass $M$
and charge $e$, and we assume our EFT to be valid below an upper energy $\Lambda_{E}$ 
(e.g. $m_{\pi}c^2 \approx 140$~MeV) in the center-of-mass frame (CoM). 
In this reference frame, the two-body retarded $(+)$ and advanced $(+)$ unperturbed Green's function operator are given by
\begin{equation}
\hat{G}_0^{(\pm)}(E) = M\int_{\mathbb{R}^3}\frac{\mathrm{d}^3 q}{(2\pi)^3}
\frac{|\mathbf{q},-\mathbf{q}\rangle\langle\mathbf{q},-\mathbf{q}|}{\mathbf{p}^2 - \mathbf{q}^2 \pm \mathrm{i}\varepsilon}~,
\label{E-2.0-01}
\end{equation}
where $\pm \mathbf{p}$ ($\pm \mathbf{q}$) are the momenta of two
incoming (outcoming) particles in the CoM frame, such that $|\mathbf{p}| = |\mathbf{q}|$ and $E = \mathbf{p}^2/M$
is the energy eigenvalue at which the Green's functions are evaluated.
The potentials are constructed in terms
of the four-fermion operators that transform as the $2\ell +1$-dimensional
irreducible representation of $\mathrm{SO(3)}$,
 \begin{equation}
V^{(\ell)}(\mathbf{p},\mathbf{q}) \equiv \langle \mathbf{q}, -\mathbf{q} | \hat{\mathcal{V}}^{(\ell)}|
\mathbf{p},
-\mathbf{p} \rangle =  \left(c_0^{(\ell)} + c_2^{(\ell)}\mathbf{p}^2  + c_4^{(\ell)}\mathbf{p}^4
+ \dots\right) (|\mathbf{p}||\mathbf{q}|)^{\ell}\mathcal{P}_{\ell}\left(\frac{\mathbf{p}\cdot\mathbf{q}}{|\mathbf{p}||\mathbf{q}|}\right)\label{E-2.0-02}
\end{equation}
where $\mathcal{P}_{\ell}$ is a Legendre polynomial, $\hat{\mathcal{V}}^{(\ell)}$
is the potential operator and the  $c_{2j}^{(\ell)}$ are low-energy constants (LECs). 
As shown in ref.~\cite{BeS14}, the polynomials in the first round bracket of eq.~\eqref{E-2.0-02}, can be
encoded by a single vertex with energy-dependent coefficient $C(E^*)$ ($D(E^*)$ and $F(E^*)$)
for S-waves (P- and D-waves), where $E^*$ represents the CoM energy of the particles. Hence, 
the Lagrangian density for fermions coupled to one unit of angular momentum assumes the form
\begin{equation}
\mathcal{L} = \psi^{\dagger}\left[\mathrm{i}\hbar\partial_t + \frac{\hbar^2\nabla^2}{2M}\right]\psi
+ \frac{D(E^*)}{8}(\psi \overleftrightarrow{\nabla}_i \psi)^{\dagger}(\psi \overleftrightarrow{\nabla}_i \psi)~,
\label{E-2.0-03}
\end{equation} 
where $\overleftrightarrow{\nabla} = \overleftarrow{\nabla} - \overrightarrow{\nabla}$ denotes the Galilean
invariant derivative. Recalling the Feynman rules in app.~A of ref.~\cite{StM21},
the two-body $\ell=1$ potential in momentum space becomes
\begin{equation}
V^{(1)}(\mathbf{p},\mathbf{q}) \equiv \langle \mathbf{q}, -\mathbf{q} \lvert \hat{\mathcal{V}}^{(1)}
 \lvert \mathbf{p}, -\mathbf{p}\rangle = D(E^*)~\mathbf{p}\cdot \mathbf{q}~.\label{E-2.0-04}
\end{equation}

\begin{figure} [hb]
\includegraphics[width=0.49\columnwidth]{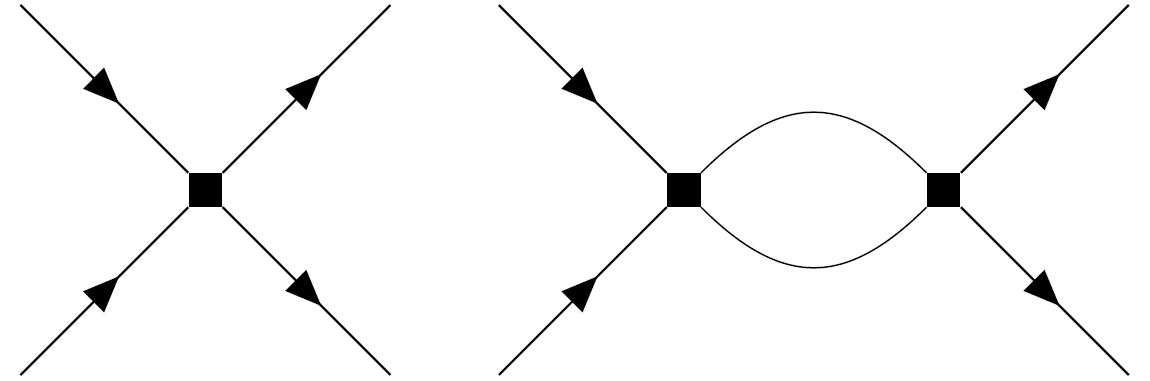}
\includegraphics[width=0.48\columnwidth]{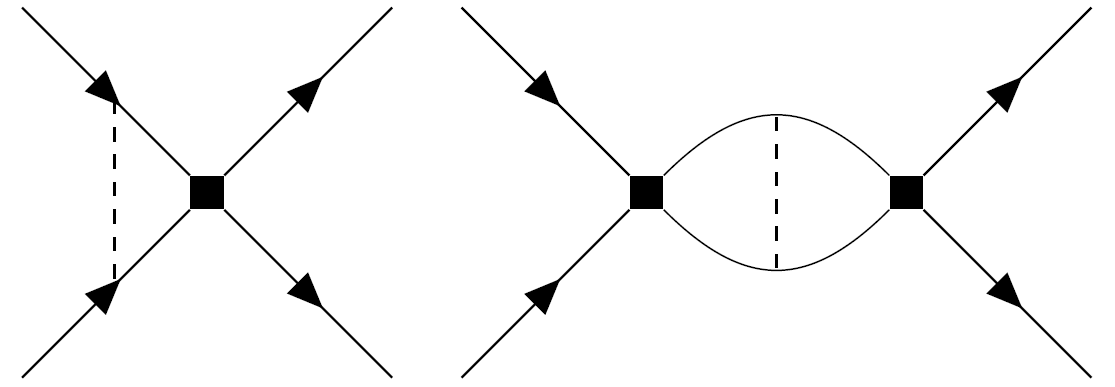}
\caption{Tree-level and 1-loop (\textit{bubble}) diagrams
representing fermion-fermion elastic scattering with the strong $\ell = 1$ potential in eq.~\eqref{E-2.0-04}
alone (left) and with scalar-photon insertions (right), see eq.~\eqref{E-2.1-02}.}
\label{F-2-01}
\end{figure}

Irrespectively on the angular momentum content of the strong interactions, the Feynman diagrams contributing
to the T-matrix, for the two-body fermion-fermion or fermion-antifermion scattering processes assume 
the form of chains of bubbles. With reference to the strong P-wave vertex in eq.~\eqref{E-2.0-04}, the
expression for the two-body scattering amplitude can be written as,
\begin{equation}
\mathrm{i}T_{\mathrm{S}}(\mathbf{p},\mathbf{q}) = \mathrm{i} \langle \mathbf{q}, -\mathbf{q} \lvert
\left[\hat{\mathcal{V}}^{(1)}+ \hat{\mathcal{V}}^{(1)}\hat{G}_0^{E}\hat{\mathcal{V}}^{(1)} 
+ \hat{\mathcal{V}}^{(1)}\hat{G}_0^{E}\hat{\mathcal{V}}^{(1)}\hat{G}_0^{E}\hat{\mathcal{V}}^{(1)} + \ldots
\right]\lvert \mathbf{p}, -\mathbf{p}\rangle~,\label{E-2.0-05}
\end{equation}
where $\hat{G}_0^{E} \equiv \hat{G}_0^{(+)}(E)$ is the two-body unperturbed retarded Green's
function operator in eq.~\eqref{E-2.0-01}. 

As shown in sec.~2 of ref.~\cite{StM21}, the infinite superposition of 
chains of bubbles translates into a geometric series of ratio $D(E^*)\mathbb{J}_{0}$ where
\begin{equation}
(\mathbb{J}_0)_{ij}^{\mathrm{PDS}} = \partial_{i}\partial_{j}'G_0^E(\mathbf{r},\mathbf{r}')\Big\lvert_{\substack{\mathbf{r}
= \mathbf{0} \\\mathbf{r}'= \mathbf{0}}}^{\mathrm{PDS}}   = -\delta_{ij}\frac{M}{4\pi}
\left(\frac{\mathrm{i}|\mathbf{p}|^3}{3} +\mu \frac{\mathbf{p}^2}{2}\right)~.\label{E-2.0-06}
\end{equation}
and is evaluated in dimensional regularization with renormalization mass $\mu$ in the PDS scheme. Therefore,
the strong scattering amplitude can be recast in the form
\begin{equation}
T_{\mathrm{S}}(\mathbf{p},\mathbf{q}) = 
\mathbf{q}\cdot\frac{ D(E^*)}{\mathbbm{1}-D(E^*)\mathbb{J}_{0}}\mathbf{p}
= \frac{12\pi}{M}\frac{D(E^*)\mathbf{q}\cdot\mathbf{p}}{\frac{12\pi}{M}
+\mathrm{i}D(E^*)|\mathbf{p}|^3}
~.\label{E-2.0-07}
\end{equation}
 which can be compared with eq.~(2) in ref.~\cite{KoR00} for the S-wave interactions.
 Recalling the effective-range expansion (ERE) \cite{Omn70} and the 
$\ell = 1$ contribution to the partial wave expansion (PWE) of the scattering ampltiude, the expression for the scattering 
length in eq.~(24) in ref.~\cite{StM21} can be derived, whereas the effective range parameter 
$r_0^{(1)}$ vanishes.
Exploiting the PDS expression of $\mathbb{J}_{0}$ in eq.~\eqref{E-2.0-06}, 
the renormalized coupling constant $D(E^*,\mu)$ is obtained
\begin{equation}
D(E^*,\mu) = \frac{12\pi}{M}\left(\frac{1}{a^{(1)}}-\mu\frac{\mathbf{p}^2}{2}\right)^{-1}~.\label{E-2.0-08}
\end{equation}
where $a^{(1)}$ denotes the scattering length (cf. eq.~(7) of ref.~\cite{KoR00} for the S-wave case counterpart).


\subsection{Coulomb corrections}\label{S-2.1}

We introduce the electromagnetic interactions in the framework of non-relativistic quantum electrodynamics
(NRQED) as in refs.~\cite{CaL86}. Since transverse photons couple proportionally to the fermion
momenta, the Coulomb photons dominate at low energies \cite{KoR00}. Therefore, we choose to retain 
in the Lagrangian only the scalar photon field, $\phi$, and its lowest order coupling to the matter fields \cite{CaL86}.
As a consequence, the Lagrangian density of the system in eq.~\eqref{E-2.0-03} picks up the following contribution,
\begin{equation}
\mathcal{L}^{\mathrm{NRQED}~\mathrm{corr}} = -\frac{1}{2} \nabla\phi \cdot \nabla\phi - e\phi~\psi^{\dagger}\psi~.
\label{E-2.1-01}
\end{equation}
The added interaction corresponds to the Coulomb force, that in momentum space reads 
\begin{equation}
V_{\mathrm{C}} (\mathbf{p},\mathbf{q}) \equiv \langle \mathbf{q}, -\mathbf{q}|\hat{V}_{\mathrm{C}}|\mathbf{p},-\mathbf{p}\rangle
= \frac{e^2}{(\mathbf{p}-\mathbf{q})^2 +\lambda^2}~,\label{E-2.1-02}
\end{equation}
where $\lambda$  is an IR regulator. Consequently, the T-matrix is enriched by new classes of diagrams (cf.
fig.~\ref{F-2-01}), in which the Colulomb photon insertions (\textit{ladders}) appear, either between the external legs and within the loops. 
Unlike transverse photons, the scalar ones do not propagate between different bubbles.
In low-momentum sector $\eta = \alpha M /2 |\mathbf{p}|^2 \gg 1$, thus
the ladders have to be incorporated in the amplitude of two-body processes to all orders in $\alpha$.
This amounts to replacing the free propagators, $G_0^{(\pm)}(\mathbf{p},\mathbf{q})$, in the loops on the left of fig.~\ref{F-2-01}
with the Coulomb propagators, that in operator form read
\begin{equation}
\hat{G}_{\mathrm{C}}^{(\pm)} = M\int_{\mathbb{R}^3}\frac{\mathrm{d}^3\mathbf{q}}{(2\pi)^3}
\frac{|\psi_{\mathbf{q}}^{(\pm)}\rangle\langle\psi_{\mathbf{q}}^{(\pm)}|}{\mathbf{p}^2-\mathbf{q}^2\pm \mathrm{i}\varepsilon}~.
\label{E-2.1-03}
\end{equation}
where $\psi_{\mathbf{q}}^{(\pm)}$ are outgoing ($+$) and incoming ($-$) Coulomb spherical waves, see eqs.~(39) and 
(40) in ref.~\cite{StM21}, whose squared modulus evaluated at $\mathbf{r} = 0$
coincides with the \textit{Sommerfeld factor} \cite{HuB59}, 
\begin{equation}
C_{\eta}^2 \equiv |\psi_{\mathbf{p}}^{(\pm)}(0)|^2 = e^{-\pi\eta}\Gamma(1+\mathrm{i}\eta)\Gamma(1-\mathrm{i}\eta)
= \frac{2\pi\eta}{e^{2\pi\eta}-1}~.\label{E-2.1-04}
\end{equation}
Moreover, thanks to the self-consistent Dyson-like identity in eq. (45) of ref.~\cite{StM21} the eigenstates of the complete system, 
$|\chi_{\mathbf{p}}^{(\pm)}\rangle$, with $\hat{V}_{\mathrm{S}} \equiv \hat{\mathcal{V}}^{(1)}$ can be rewritten in terms of the Coulomb states,
\begin{equation}
|\chi_{\mathbf{p}}^{(\pm)}\rangle = \left[1+\sum_{n=1}^{+\infty}(\hat{G}_{\mathrm{C}}^{(\pm)}\hat{V}_{\mathrm{S}})^n\right]|
\psi_{\mathbf{p}}^{(\pm)}\rangle~.\label{E-2.1-05}
\end{equation}
Second, the T-matrix $T_{\mathrm{S}}$ is now replaced by the superposition of the purely electrostatic scattering amplitude
$T_{\mathrm{C}}(\mathbf{p}',\mathbf{p}) = \langle \mathbf{p}'|\hat{V}_{\mathrm{C}}| \psi_{\mathbf{p}}^{(+)}\rangle$ and the
 strong scattering amplitude modified by Coulomb corrections, $T_{\mathrm{SC}}(\mathbf{p}',\mathbf{p})
= \langle \psi_{\mathbf{p}'}^{(-)}|\hat{V}_{\mathrm{S}}|\chi_{\mathbf{p}}^{(+)}\rangle$. 
Both the latter admit an expansion in in terms of the Legendre polynomials (cf. eqs.~(48) and (49) in ref.~\cite{StM21}), in which
the \textit{Coulomb phase shift}, $\sigma_1 = \arg \Gamma(2 + \mathrm{i}\eta)$, is introduced besides $\delta_1$. Analogously, the P-wave ERE is 
of the generalized counterpart for the (repulsive) Coulomb interaction (cf.~\cite{KMB82}),
\begin{equation}
\mathbf{p}^2\left(1+\eta^2\right)\left[C_{\eta}^2|\mathbf{p}|(\cot\delta_1-\mathrm{i}) + \alpha M H(\eta)\right] 
= -\frac{1}{a_{\mathrm{C}}^{(1)}} +\frac{1}{2}r_0^{(1)}\mathbf{p}^2 + r_1^{(1)}\mathbf{p}^4 + \ldots~,\label{E-2.1-06}
\end{equation}
where $a_{\mathrm{C}}^{(1)}$, $r_0^{(1)}$ and $r_1^{(1)}$ are the scattering length, the effective range and the shape parameter,
respectively and $H(\eta)$ is a function of the parameter $\eta$ alone and is defined in eq.~(52) of ref.~\cite{StM21}.


\subsection{Repulsive channel}\label{S-2.2}

A this stage, the Coulomb-corrected strong amplitude, $T_{\mathrm{SC}}(\mathbf{p},\mathbf{p}')$,
for the fermion-fermion scattering process can be computed. The crucial tool for the derivation, outlined in sec.~2.2 and app.~C
of ref.~\cite{StM21} in full detail, is represented by eq.~\eqref{E-2.1-05}, which allows for the evaluation of the 
T-matrix element in terms of the free Coulomb eigenstates alone and for the factorization of the contributions 
of each P-wave vertex in the amplitude associated to each bubble diagram. It follows that 
$T_{\mathrm{SC}}(\mathbf{p},\mathbf{p}')$ can be again expressed as the sum of a geometric series,
\begin{equation}
\mathrm{i}T_{\mathrm{SC}}(\mathbf{p}',\mathbf{p}) =\mathrm{i}D(E^*)~\nabla'\psi_{\mathbf{p}'}^{(-)*}(\mathbf{r}') \Big
\lvert_{\mathbf{r}'=0} \cdot \left[\mathbbm{1}  + D(E^*)\mathbb{J}_{\mathrm{C}} + D(E^*)^2\mathbb{J}_{\mathrm{C}}^2
 +\ldots  \right] \nabla\psi_{\mathbf{p}}^{(+)}(\mathbf{r}) \Big\lvert_{\mathbf{r}=0}~,
\label{E-2.2-01}
\end{equation}
whose ratio is given by the Hessian matrix of the two-body Coulomb propagator, $\mathbb{J}_{\mathrm{C}}$, multiplied 
by $D(E^*)$. Since the nondiagonal matrix elements vanish in DR (cf. eq.~(4.3.14) in ref.~\cite{Col84}),
\begin{equation}
(\mathbb{J}_{\mathrm{C}})_{ij} =  \partial_{i}\partial_{j}'G_{\mathrm{C}}^{(+)}(\mathbf{r},\mathbf{r}')|_{\mathbf{r},\mathbf{r}'=0}  
= \lim_{d \rightarrow 3} M\frac{\delta_{ij}}{d}\int_{\mathbb{R}^d} \frac{\mathrm{d}^ds}{(2\pi)^d}
\frac{2\pi \eta(s)~s^2}{e^{2\pi\eta(s)}-1}  \frac{1+\eta(s)^2}{\mathbf{p}^2-\mathbf{s}^2+\mathrm{i}\varepsilon}
\equiv  \mathbbm{j}_{\mathrm{C}}\delta_{ij}~,\label{E-2.2-02}
\end{equation}
the Coulomb-corrected strong scattering amplitude, $T_{\mathrm{SC}}(\mathbf{p},\mathbf{p}')$, can be further simplified as 
\begin{equation}
T_{\mathrm{SC}}(\mathbf{p}',\mathbf{p}) = (1+\eta^2) C_{\eta}^2 \frac{D(E^*) ~e^{2\mathrm{i}\sigma_1} \mathbf{p}\cdot\mathbf{p}'}
{1-D(E^*)~\mathbbm{j}_{\mathrm{C}}}~.\label{E-2.2-03}
\end{equation} 
Evaluating the diagonal matrix elements $\mathbbm{j}_{\mathrm{C}}$ in dimensional regularization (cf. eq.~(89) in ref.~\cite{StM21}) with the PDS scheme
and exploiting the $\ell = 1$ component of the partial wave expansion for $T_{\mathrm{SC}}(\mathbf{p}',\mathbf{p})$ in eq.~(49) of ref.~\cite{StM21},
an expression for $\cot \delta_1$ in terms of the cupling constants $\alpha$ and $D(E^*)$ can be
obtained. Finally, equipped with the latter formula and the generalized ERE in eq.~\eqref{E-2.1-06}, the scattering parameters can be determined
in closed form. In particular, the momentum-independent contributions to $\mathbbm{j}_{\mathrm{C}}$, yield the expression for the scattering length,
\begin{equation}
\frac{1}{a_{\mathrm{C}}^{(1)}} = \frac{12\pi}{M D(E^*)} + \frac{\alpha^2M^2\mu}{8}\left(\pi^2-3\right) 
- \frac{\alpha^3 M^3}{4}\left[ \frac{1}{3-d} +\zeta(3) -\frac{3}{2}\gamma_E 
+ \frac{4}{3} +\log \frac{\mu\sqrt{\pi}}{\alpha M}\right] ~.\label{E-2.2-04}
\end{equation}
where $\zeta(3)$ and $\gamma_E$ are the Apéry and Euler-Mascheroni constants, respectively. Besides, grouping the quadratic terms in $\mathbf{p}$, a 
\textit{purely Coulomb} expression for the effective range is recovered, 
\begin{equation}
r_0^{(1)} = \alpha M \left[ \frac{2}{3-d} + \frac{8}{3} -3\gamma_E + 2\log\frac{\mu\sqrt{\pi}}{\alpha M} \right] -3\mu~.
\label{E-2.2-05} 
\end{equation}


\section{The finite-volume environment}\label{S-3.0}

We transpose the physical system into a cubic box with edges of length $L$ and we continue
 analythically fields and wavefunctions outside the box via periodic boundary conditions (PBCs). 
Consequently, a free particle subject to PBCs carries a momentum $\mathbf{p}=2\pi\mathbf{n}/L$,
with $\mathbf{n} \in \mathbb{Z}^3$.\\
Furthermore, in this environment the Ampère and 
Gauss laws are violated \cite{BeS14}. The issue is circumvented by adding a uniform background charge density, 
a procedure that corresponds to the removal of Coulomb-photon propagators with the zero exchanged momentum
 \cite{BeS14}. Discarding the latter, the particle's momentum is restricted to 
$|\mathbf{p}| \geq 2\pi/L$, whereas the viability of the perturbation treatment of QED 
implies $\eta = \alpha M /2|\mathbf{p}| < 1$. Combining the two constraints, it follows that the 
photon field insertions can be treated perturbatively if $ML \ll 1/\alpha$ \cite{BeS14}. The condition is realized
when the volume is sufficiently large, as in present-day Lattice QCD calculations \cite{BeS14}. \\
In addition, the masses of spinless composite particles with unit charge receive LO finite-volume corrections 
proportional to the inverse of L (cf. eqs.~(6) and (19) of ref.~\cite{DaS14}),
\begin{equation}
\Delta M \equiv M^L - M = \frac{\alpha}{2\pi L}\left[\sum_{\mathbf{n}\neq 0}^{\Lambda_n}\frac{1}{|\mathbf{n}|^2}
- 4\pi\Lambda_n\right] + \mathcal{O}\left(\alpha^2; \textstyle{\frac{\alpha}{L^2}}\right)~,\label{E-3.0-01}
\end{equation}
where the sum of the 3D Riemann series regulated by the spherical cutoff $\Lambda_n$ is denoted
with $\mathcal{I}^{(0)} \approx-8.913632$ (cf. app.~D.1 in ref.~\cite{StM21}) and the $L$ superscript denotes henceforth 
FV quantities. Reabsorbing these shifts through the definition of \textit{primed} scattering parameters (cf. eqs.~(134)-(138) 
in ref.~\cite{StM21}), the finite-volume ERE takes the form
\begin{equation}
\mathbf{p}^2 (1+\eta^2)[C_{\eta}^2|\mathbf{p}|(\cot\delta_1^L-\mathrm{i}) + \alpha M H(\eta)] 
= -\frac{1}{{a'_{\mathrm{C}}}^{(1)}}+\frac{1}{2}{r'_0}^{(1)}\mathbf{p}^2  + {r'_1}^{(1)}\mathbf{p}^4  + \ldots~.
\label{E-3.0-02}
\end{equation}


\subsection{Quantization condition}\label{S-3.1}

We derive the conditions that determine the
counterpart of the $\ell=1$ energy eigenvalues in the cubic region. These states transform as the three-dimensional
irreducible representation $T_1$ (in Sch\"onflies's notation) of the cubic rotation group \cite{SEM18}. 
The former correspond with the singularities of the two-point correlation
function $G_{\mathrm{SC}}^{(\pm)}(\mathbf{r},\mathbf{r}')$, whose expression can be derived in closed from the self-consistent 
equation connecting $G_{\mathrm{SC}}^{(\pm)}$ and $G_{\mathrm{C}}^{(\pm)}$ (cf. eq.~(144) in ref.~\cite{StM21}).\\
In finite volume $G_{\mathrm{SC}}^{(\pm)}$ $\mapsto$ $G_{\mathrm{SC}}^{(\pm), L}$ and the quantization condition becomes
 \begin{equation}
\frac{\mathbbm{1}}{D^L(E^*)} = \nabla_{\mathbf{r}_i} \otimes \nabla_{\mathbf{r}_{i+1}} G_{\mathrm{C}}^{(+),L}(\mathbf{r}_i,
\mathbf{r}_{i+1}) \Big\lvert_{\substack{\mathbf{r}_i =\mathbf{0} \\ \mathbf{r}_{i+1}=\mathbf{0}}} \equiv 
\mathbbm{J}_{\mathrm{C}}^L(\mathbf{p})~.\label{E-3.1-01}
\end{equation}
As noticed in sec.~\ref{S-3.0}, we are allowed to expand $\mathbbm{j}_{\mathrm{C}}^L(\mathbf{p})$ 
in powers of the fine-structure constant and truncate the series to order $\alpha$. With the aim of regulating the 3D Riemann sums 
appearing in the expression for $\mathbbm{J}_{\mathrm{C}}^L(\mathbf{p})$ in eq.~(148) in ref.~\cite{StM21} while maintaining 
the mass-independent renormalization scheme (cf. sec.~II B of ref.~\cite{BeS14}), the FV quantization conditions can 
be recast as 
\begin{equation}
\frac{\mathbbm{1}}{D^L(E^*)} - \mathfrak{Re} \mathbb{J}_{\mathrm{C}}^{\{\mathrm{DR}\}} (\mathbf{p})
= \mathbb{J}_{\mathrm{C}}^L(\mathbf{p}) - \mathfrak{Re} \mathbb{J}_{\mathrm{C}}^{\{\Lambda\}} (\mathbf{p})~,\label{E-3.1-02}
\end{equation}
where $\mathbb{J}_{\mathrm{C}}^{\{\Lambda\}} (\mathbf{p})$ and $\mathbb{J}_{\mathrm{C}}^{\{\mathrm{DR}\}} (\mathbf{p})$
denote the $\mathcal{O}(\alpha)$ approximations of $\mathbb{J}_{\mathrm{C}}$ computed in the cutoff- and dimensional
regularization (DR) schemes in eqs.~(165) and (174) of ref.~\cite{StM21} respectively. A scalar counterpart for eq.~\eqref{E-3.1-02} can be 
recovered by taking the trace of the matrices. \\
At this stage, by exploiting eq.~\eqref{E-2.2-03}, the expression for $\mathbbm{j}_{\mathrm{C}}$ in DR
to all orders in $\alpha$ given in eq.~(89) of ref.~\cite{StM21} and eq.~(51) in ref.~\cite{StM21}, an infinite-volume relation 
between the P-wave phase shift and the strong coupling constant $D(E^*)$ can be obtained, see eq.~(177) in ref.~\cite{StM21}.
The latter can be immediately fitted to the cubic FV case, yielding 
\begin{equation}
\mathbf{p}^2(1+\eta^2)[C_{\eta}^2|\mathbf{p}|(\cot\delta_1^L-\mathrm{i}) + \alpha M H(\eta)] 
= -\frac{12\pi}{M D^L(E^*)} + \alpha M \mathbf{p}^2\left[\frac{4}{3} - \frac{3}{2}\gamma_E
+ \log\left(\frac{\mu\sqrt{\pi}}{\alpha M}\right) \right]~.\label{E-3.1-03}
\end{equation}
Then, exploiting the scalar counterpart of eq.~\eqref{E-3.1-01}, the coupling constant $D^L(E^*)$ in eq.~\eqref{E-3.1-03} can 
be replaced by the regulated FV expression for $\mathbbm{j}_{\mathrm{C}}$. Finally, expanding the l.h.s. of eq.~\eqref{E-3.1-03}
in power series of the squared momentum, the finite-volume ERE in eq.~\eqref{E-3.0-02} becomes
\begin{equation}
\begin{gathered}
-\frac{1}{{a'_C}^{(1)}} + \frac{1}{2}{r'_0}^{(1)} \mathbf{p}^2 + {r'_1}^{(1)} \mathbf{p}^4 +  {r'_2}^{(1)} \mathbf{p}^6 +  {r'_3}^{(1)} \mathbf{p}^8 +  ... \\
 =  \frac{4\pi}{L^3}\mathcal{S}_0(\tilde{\mathbf{p}}) + \frac{\mathbf{p}^2}{\pi L}\mathcal{S}_1(\tilde{\mathbf{p}})  
- \frac{\alpha M \mathbf{p}^2}{4\pi^4} \mathcal{S}_2(\tilde{\mathbf{p}})  -\frac{\alpha M}{\pi^2 L^2}\mathcal{S}_3(\tilde{\mathbf{p}}) 
+ ... + \alpha M\mathbf{p}^2 \left[ \log\left(\frac{4\pi}{\alpha M L}\right)-\gamma_E \right]~,\label{E-3.1-04}
\end{gathered}
\end{equation}
where the $\mathcal{S}_n(\tilde{\mathbf{p}})$ with $n \leq 3$ denote the \textit{L\"uscher functions}, given by
\begin{equation}
\mathcal{S}_0(\tilde{\mathbf{p}}) = \sum_{\mathbf{n}}^{\Lambda_n}1 - \frac{4\pi}{3}\Lambda_n^3~, \hspace{0.5cm}
\mathcal{S}_1(\tilde{\mathbf{p}}) = \sum_{\mathbf{n}}^{\Lambda_n}\frac{1}{\mathbf{n}^2-\tilde{\mathbf{p}}^2}
- 4\pi\Lambda_n~,\label{E-3.1-05}
\end{equation}
\begin{equation}
\mathcal{S}_2(\tilde{\mathbf{p}}) = \sum_{\mathbf{n}}^{\Lambda_n}\sum_{\mathbf{m}\neq \mathbf{n}}^{\infty} \frac{1}{\mathbf{n}^2
-\tilde{\mathbf{p}}^2} \frac{1}{\mathbf{m}^2-\tilde{\mathbf{p}}^2}\frac{1}{|\mathbf{n}-\mathbf{m}|^2} - 4\pi^4 \log
\Lambda_n~,\label{E-3.1-06}
\end{equation}
and
\begin{equation}
\mathcal{S}_3(\tilde{\mathbf{p}}) = \sum_{\mathbf{n}}^{\Lambda_n}\sum_{\mathbf{m}\neq \mathbf{n}}^{\infty} \frac{1}{\mathbf{n}^2
-\tilde{\mathbf{p}}^2} \frac{1}{\mathbf{m}^2-\tilde{\mathbf{p}}^2}\frac{\mathbf{m}\cdot \mathbf{n}
- \tilde{\mathbf{p}}^2}{|\mathbf{n}-\mathbf{m}|^2} - \pi^4 \Lambda_n^2~.\label{E-3.1-07}
\end{equation}


\subsection{Approximate energy eigenvalues}\label{S-3.2}

The $\mathcal{O}(\alpha)$ ERE presented in eq.~\eqref{E-3.1-04} acquires a pivotal role in the derivation
of the approximate expressions in the finite volume for the lowest bound and unbound eigenvalues of the Hamiltonian two-fermion 
system given by $\hat{H}_0 + \hat{V}_{\mathrm{C}} + \mathcal{V}^{(1)}$ where $\hat{H}_0$ denotes the kinetic term associated to the two particles
in the CoM frame. Under the hypotesis that $L$ is sufficiently large, the expansions become perturbative both in $\alpha$ and
 in $1/L$ times suitable powers of the scattering parameters. 


\subsubsection{The lowest unbound state}\label{S-3.2.1}

Since $\ell = 1$ irreducible representation (irrep) of SO(3) is mapped to the $T_1$ irrep of the cubic group, 
the FV eigenstates are expected to be three-fold degenerate. Hence, the lowest
unbound state corresponds to an energy of $4\pi^2/ML^2$, associated to the 
dimensionless momentum $\tilde{p} = 1$ where $|\mathbf{p}| =  2\pi \tilde{p}/L$.

The detailed procedure for the derivation of the FV energy eigenvalue is outlined in sec.~3.3.1
of ref.~\cite{StM21} and is based on the expansion of the Lüscher functions in the ERE in eq.~\eqref{E-3.1-07}
around $\delta\tilde{p}^2 = \tilde{p}-1$. 
Once the ERE is rewritten in powers of $\delta\tilde{p}^2$, the 
equation is solved iteratively for $\tilde{p}^2$, truncating the expansion at a given power of $\tilde{p}^2$ and exploiting the outcoming 
expression in the subsequent step, which contains higher-order terms in $\tilde{p}^2$. The final expression for $E_{\mathrm{S}}^{(1,T_1)}$
is given in eq.~(218) of ref.~\cite{StM21} and reported here in concise form,
\begin{equation}
\begin{gathered}
E_{\mathrm{S}}^{(1,T_1)}  =  \frac{4 \pi^2}{M L^2}  +  \frac{4 \pi^2\delta\tilde{p}^2}{M L^2} = \frac{4 \pi^2}{M L^2} 
+ 6 \xi \frac{4\pi {a'_{\mathrm{C}}}^{(1)}}{ML^3}\left[1 + \xi^2 {a'_{\mathrm{C}}}^{(1)} {r_{0}'}^{(1)}  
- \xi \left(\frac{{a'_{\mathrm{C}}}^{(1)}}{\pi L}\right)(\mathcal{I}^{(1)} - 6)   + \ldots  \right] \\
 + \xi \frac{\alpha~{a'_{\mathrm{C}}}^{(1)}}{L^2\pi^2}\Big\{ -(2\chi_1+2\Qoppa_0) 
\left[1+ \xi^2 {a'_{\mathrm{C}}}^{(1)} {r_{0}'}^{(1)} + \xi^3 {a'_{\mathrm{C}}}^{(1)} {r_{1}'}^{(1)}\right] \\
 + \xi\left(\frac{{a'_{\mathrm{C}}}^{(1)}}{\pi L}\right)\left[
(\mathcal{I}^{(1)} - 6)(2\Qoppa_0 + 2\chi_1) + 6(\Qoppa_1-\chi_1 -2\chi_2+\tilde{\mathcal{R}}^{(1)} + 6 \mathcal{J}^{(1)})
\right] + \ldots \Big\}  + \ldots ~\label{E-3.2.1-01}
\end{gathered}
\end{equation}
where the ellipsis stands for terms of higher powers of $1/L$ and the scattering parameters and $\mathcal{I}^{(1)}$,
$\mathcal{J}^{(1)}$, $\tilde{\mathcal{R}}^{(1)}$, $\chi_i$ and $\Qoppa_i$ are sums of 3D Riemann series, 
evaluated numerically in app.~D of ref.~\cite{StM21}.

\subsubsection{The lowest bound state}\label{S-3.2.2}

To derive the lowest bound FV energy eigenvalue, we switch to imaginary
 momenta, $\mathbf{p} = \mathrm{i}\boldsymbol{\kappa}$ and evaluate the Lüscher sums in eqs.~\eqref{E-3.1-05}-\eqref{E-3.1-07}
in the limit of large dimensionless binding momentum, $\tilde{\kappa} \gg 1$. The results, derived in app.~E of ref.~\cite{StM21} in 
detail, permit to rewrite the ERE in eq.~\eqref{E-3.1-04} in terms of powers of 
$\kappa \equiv |\boldsymbol{\kappa}| = \kappa_0 + \kappa_1 + \ldots$ where $\kappa_i$ encodes the contributions to order $i$
 in $\alpha$. Discarding the higher-order terms in the scattering parameters and $1/L$ (cf. sec.~3.3.2 of ref.~\cite{StM21}),
an expression for $\kappa_1$ is recovered from the ERE, yielding the approximate eigenvalue 
\begin{equation}
E_{B}^{(1,T_1)}(L) = \frac{\kappa_0^2}{M} + \frac{2 \alpha \kappa_0^3}{3\kappa_0^2 - r_0^{(1)} \kappa_0} \left[  \log
\left( \frac{4\kappa_0}{\alpha M} \right) - \gamma_E + \frac{1}{2} \right] + \frac{\alpha\mathcal{I}^{(0)}}{\pi L}
- \frac{\alpha}{\pi^3 L^3} \frac{2\pi^4}{\kappa_0^2} \frac{1}{3k_0 - r_0^{(1)}}~.\label{E-3.2.2-01}
\end{equation}
We can infer that the leading FV contributions are positive, analogously to the
LO mass shifts for $\ell=1$ states of two-body systems with strong interactions alone in eq.~(53) of
ref.~\cite{KLH12}. Finally, the sign of the LO P-wave FV shift is opposite with respect to the S-wave one in eq.~(46) in ref.~\cite{BeS14},
\begin{equation}
\Delta E_{B}^{(1,T_1)} \equiv E_{B}^{(1,T_1)}(\infty) -  E_{B}^{(1,T_1)}(L)  \stackrel{\text{LO}}{=}  - \frac{\alpha\mathcal{I}^{(0)}}{\pi L}
\stackrel{\text{LO}}{=}  -\Delta E_{B}^{(0,A_1)} ~.\label{E-3.2.2-02}
\end{equation}

\section{Beyond P-wave interactions}\label{S-4.0}

Among the possible improvements and generalizations of the present work, outlined in sec.~4 of ref.~\cite{StM21}, we here 
concentrate on the treatment of fermion fields with coupling to two units of angular momentum. In the belief that the procedure is instrumental 
for the complete extension of the analysis in ref.~\cite{StM21} in this direction, we derive the fermion-fermion scattering amplitude, in 
which the particles are solely subject to the short range D-wave potentials.\\
Consistently with requirements in sec.~\ref{S-2.0}, the Lagrangian density of the system reads
\begin{equation}
\mathcal{L} = \psi^{\dagger}\left[\mathrm{i}\hbar\partial_t - \frac{\hbar^2\nabla^2}{2M}\right]\psi
+ \frac{F(E^*)}{64} (\psi \overleftrightarrow{\nabla}^2 \psi)^{\dagger}(\psi \overleftrightarrow{\nabla}^2 \psi)  - 3\frac{F(E^*)}{64}(\psi \overleftrightarrow{\partial}_i
\overleftrightarrow{\partial}_j\psi)^{\dagger}(\psi \overleftrightarrow{\partial}_i
\overleftrightarrow{\partial}_j \psi) ~.\label{E-4.0-01}
\end{equation} 
In momentum space, the interaction operator, $\hat{\mathcal{V}}^{(2)}$, now depends on the square of the scalar product
between the momentum of the incoming ($\pm \mathbf{p}$) and outcoming ($\pm\mathbf{q}$) particles in the CoM frame,
\begin{equation}
V^{(2)}(\mathbf{p},\mathbf{q}) \equiv \langle \mathbf{q}, -\mathbf{q} \lvert \hat{\mathcal{V}}^{(2)}
 \lvert \mathbf{p}, -\mathbf{p}\rangle = F(E^*)~\left[3(\mathbf{p}\cdot \mathbf{q})^2 - \mathbf{p}^2 \mathbf{q}^2\right]~.\label{E-4.0-02}
\end{equation}
By virtue of the self-consistent identity between the free and the 'strong' Green's functions (cf. eq.~(1.9) in ref.~\cite{Ste20}),
the T-matrix element, $T_{\mathrm{S}}(\mathbf{p},\mathbf{q})$, can be written again as a superposition of the bubble-diagram amplitudes,
\begin{equation}
\mathrm{i}T_{\mathrm{S}}(\mathbf{p},\mathbf{q}) = \mathrm{i} \langle \mathbf{q}, -\mathbf{q} \lvert
\left[\hat{\mathcal{V}}^{(2)}+ \hat{\mathcal{V}}^{(2)}\hat{G}_0^{E}\hat{\mathcal{V}}^{(2)} 
+ \hat{\mathcal{V}}^{(2)}\hat{G}_0^{E}\hat{\mathcal{V}}^{(2)}\hat{G}_0^{E}\hat{\mathcal{V}}^{(2)} + \ldots
\right]\lvert \mathbf{p}, -\mathbf{p}\rangle~.\label{E-4.0-03}
\end{equation}
see eq.~\eqref{E-2.0-05}. The explicit evaluation of lowest contributions to the two-body scattering amplitude, permits to conclude that the 
ratio of the geometric series is now given by $F(E^*)$ times the rank-four tensor $\mathbb{J}_0^{[2]}$, whose elements 
evaluated in DR with the PDS regularization scheme and renormalization mass $\mu$ are given by
\begin{equation}
(\mathbb{J}_0^{[2]})_{ijkl}(\mathbf{p}) = M\int_{\mathbb{R}^3}\frac{\mathrm{d}^3p'}{(2\pi)^3} 
\frac{p'_ip'_jp'_kp'_l}{\mathbf{p}^2-{\mathbf{p}'}^2 +\mathrm{i}\varepsilon}
\stackrel{\mathrm{PDS}}{=} -\frac{M}{4\pi} \left[\delta_{ij}\delta_{kl} + \delta_{ki}\delta_{jl} + \delta_{kj}\delta_{il}\right] 
\left( \frac{\mathrm{i} \mathbf{p}^5 }{15} + \mu \frac{\mathbf{p}^4}{8} \right)~\label{E-4.0-04}
\end{equation}
where the multiplication rule is established by the product between the tensors with four indices, 
s.t. $(R.S)_{ijmn} = R_{ijkl}S_{klmn}$. Succintly written, the resulting T-matrix becomes
\begin{equation}
T_{\mathrm{S}}(\mathbf{p},\mathbf{q}) = q_i q_j 
\left(\frac{3 F(E^*)}{\mathbbm{1}-F(E^*)\mathbb{J}_0^{[2]}}\right)_{ijkl}\mathbb{H}_{klmn} p_np_m
~,\label{E-4.0-05}
\end{equation}
where the $\mathbb{H}$ tensor reflects the Legendre polynomial underlying the D-wave interaction in eq.~\ref{E-4.0-02},
\begin{equation}
\mathbb{H}_{ijmn} = \frac{1}{3}\left[ -\delta_{ij}\delta_{mn} + \frac{3}{2}(\delta_{im}\delta_{jn} + \delta_{in}\delta_{jm}) \right]
~.\label{E-4.0-06}
\end{equation}
Performing expliticitly the tensor inversion implied by eq.~\eqref{E-4.0-05} and exploiting the expression of $\mathbb{J}_0^{[2]}$ in DR 
without the PDS term, the scattering amplitude in eq.~\eqref{E-4.0-05} can be recast as
\begin{equation}
T_{\mathrm{S}}(\mathbf{p},\mathbf{q}) = F(E^*) \frac{3(\mathbf{p}\cdot\mathbf{q})^2
 - \mathbf{p}^2\mathbf{q}^2}{1-\frac{2}{5}F(E^*)\mathbbm{j}_0^{[2]}} = F(E^*) \frac{4\pi}{M}
 \frac{3(\mathbf{p}\cdot\mathbf{q})^2 - \mathbf{p}^2\mathbf{q}^2}{\frac{4\pi}{M} 
+  F(E^*)\mathrm{i}\frac{2}{5}\mathbf{p}^5}\label{E-4.0-07}
\end{equation}
where $\mathbbm{j}_0^{[2]}$ is defined as the scalar part of the tensor elements of $\mathbb{J}_0^{[2]}$.  
Making use of the ERE in ref.~\cite{Omn70} and of the $\ell = 2$ term of the PWE of $T_{\mathrm{S}}(\mathbf{p},\mathbf{q})$, 
the scattering length can be expressed in terms of the energy-dependent parameter $F(E^*)$ as 
\begin{equation}
a^{(2)} = \frac{M}{4\pi} \frac{2}{5} F(E^*)~,\label{E-4.0-07}
\end{equation}
whereas the effective range, $r_0^{(2)}$, vanishes. Equipped with the PDS expression of $\mathbb{J}_{0}^{[2]}$ in eq.~\eqref{E-4.0-04}, 
the D-wave counterpart of eq.~\eqref{E-2.0-08} is eventually obtained,
\begin{equation}
F(E^*,\mu) = \frac{4\pi}{M} \frac{5}{2}\left(\frac{1}{a^{(2)}} - \frac{15}{8}\mu \mathbf{p}^4\right)^{-1}~.\label{E-4.0-08}
\end{equation}


\section{Conclusion}\label{S-5.0}

In this work we have recapitulated the P-wave extension of the investigation on QED effects 
in low-energy fermion-fermion scattering shown in ref.~\cite{StM21} in the framework of pionless EFT.
The comprehensive formalism in ref.~\cite{KoR00} based on the full Coulomb propagator permitted us to
derive expressions of the scattering amplitude in infinite volume that are exact to all orders in the fine-structure
constant. In closing, the derivation of the two-body scattering amplitude for the Lagrangian with D-wave interactions 
alone provides a prelude for the future extension of the present investigation in that direction.

The finite-volume results recapitulated in this paper confirm the power-law dependence on $1/L$ of the P-wave FV
energy corrections to the lowest bound and scattering eigenvalues in presence of QED, displayed
in ref.~\cite{BeS14} for S-wave states. 

Thanks to the momentum discretization, perturbative QED comes 
back into play in our system, even in the very low-energy regime.
In the final part of our analysis we find that for the $\ell=0$ and $1$ two-body bound eigenstates the sign of the shift depends directly on
the parity of the wavefunction associated to the energy state, whose tails are truncated at the boundaries of the cubic 
box, as claimed in ref.~\cite{KLH12} in presence of short range forces alone.


\section*{Acknowledgments}

We express gratitude to Ulf-G. Meißner, co-author of the main reference of this contribution. Besides, we acknowledge financial 
support from the Deutsche Forschungsgemeinschaft (Sino-German CRC 110, grant No. TRR~110)
and the VolkswagenStiftung (grant No. 93562) and the computational resources provided by RWTH Aachen (JARA 0015 project).


\end{document}